\documentclass[12pt,float,cite]{iopart}
\usepackage{graphicx}
\begin{document}
%
%
\title[Spreading with immunization in high dimensions]
      {Spreading with immunization in high dimensions}
\author{Stephan~M~Dammer$^{1}$ and Haye Hinrichsen$^{2}$}

\address{$^{1}$Institut f\"ur Physik,
             Universit{\"a}t Duisburg-Essen, D-47048 Duisburg, Germany}

\address{$^{2}$Fakult\"at f\"ur Physik und Astronomie, 
             Universit\"at W\"urzburg, D-97074 W\"urzburg, Germany}

\begin{abstract}
We investigate a model of epidemic spreading with partial immunization
which is controlled by two probabilities, namely, for first infections, $p_0$,
and reinfections, $p$. When the two probabilities are equal, the model 
reduces to directed percolation, while for perfect immunization one
obtains the general epidemic process belonging to the universality
class of dynamical percolation. We focus on the critical behavior in
the vicinity of the directed percolation point, especially in high
dimensions $d>2$. It is argued that the clusters of immune sites are compact for $d\leq 4$. 
This observation implies that a recently introduced scaling argument, suggesting a stretched
exponential decay of the survival probability for $p=p_c$, $p_0\ll p_c$ in one spatial dimension,
where $p_c$ denotes the critical threshold for directed percolation,
should apply in any dimension $d \leq 3$ and maybe for $d=4$ as well. Moreover, we show that the
phase transition line, connecting the critical points of directed
percolation and of dynamical percolation, terminates in the critical
point of directed percolation with vanishing slope for $d<4$
and with finite slope for $d\geq 4$. Furthermore, an exponent is
identified for the temporal correlation length for the case of $p=p_c$ 
and $p_0=p_c-\epsilon$, $\epsilon\ll 1$, which is different from the
exponent $\nu _\parallel$ of directed percolation. We also improve
numerical estimates of several critical parameters and exponents,
especially for dynamical percolation in $d=4,5$.
\end{abstract}

\maketitle
\def\xvec{\text{\bf x}}
\def\text#1{\mbox{#1}}
\parskip 1mm 
%
\section{Introduction}
\label{intro}
%
The study of stochastic models that describe the spreading of a
nonconserved agent is currently a very active field of research in
nonequilibrium statistical
physics~\cite{MarroDickman,Hinrichsen,OdorReview}. Fundamental interest stems
from the fact that these models exhibit continuous nonequilibrium
phase transitions from a fluctuating active phase into one or several
absorbing states. Moreover, such models are motivated by a variety of
possible applications~\cite{Brazil} such as epidemic
spreading~\cite{Mollison77,mut}, catalytic reactions~\cite{ZGB86}, or flowing
sand~\cite{HJRD99,HJRD99b}.

Usually models for epidemic spreading are defined on a lattice whose
sites can be either occupied (active, infected) or empty (inactive,
healthy). The dynamic rules involve two competing processes, namely,
spreading of activity to neighboring sites (infection) and
spontaneous decay (recovery). Depending on the relative frequency of
infection and recovery the process either has a finite probability to
survive in an infinite system or it will eventually die out. Since the
spreading agent is not allowed to be created spontaneously,  once the
empty lattice is reached, the process is trapped in a so-called
{\em absorbing state} from which it cannot escape. Usually the transition
between infinite spreading and recovery is a critical phenomenon
characterized by large-scale fluctuations associated with certain
universality classes. Therefore, the classification of all possible
types of phase transitions into absorbing states is presently one of
the major goals of nonequilibrium statistical physics.

The most prominent universality class of phase transitions into an
absorbing state is that of directed percolation (DP)~\cite{Kinzel85}, as
described by Reggeon field theory~\cite{Reggeon1,Reggeon2,Reggeon3}. Models in this
class describe short-range spreading of a non-conserved agent in a
medium without temporal memory effects. Additionally, it is assumed
that the rates for spreading and recovery are constant in space and
time (no quenched randomness).

\begin{figure}
\centerline{\includegraphics[width=110mm]{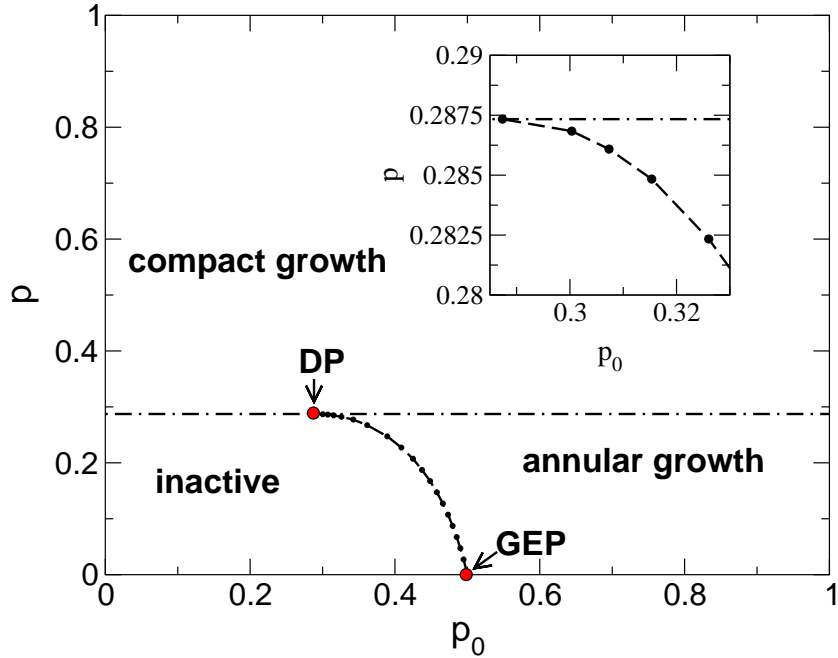}}
\caption{\label{phasediag}  Phase diagram of the epidemic process with finite
  immunization on a square lattice in 2+1 dimensions. $p_0$ denotes the primary
infection probability and $p$ the reinfection probability. The inset
  shows the vicinity of the DP point.}
\end{figure}

If the medium has a memory such that each site can be infected only
once, the transition does no longer belong to the DP class, instead
one obtains the so-called ``general epidemic process''
(GEP)~\cite{Mollison77,Grassberger83}. In the language of epidemic
processes this type of memory accounts for {\em perfect}
immunization. In contrast to DP, where sites can be reinfected without
restriction, a general epidemic process can only spread in those parts
of the system which have not been infected before.  Nevertheless, in spatial
dimensions $d \geq 2$, infinite spreading from a single
infected site in a non-immune environment is still possible, provided
that the susceptibility to primary infections is sufficiently
large. In this case activity propagates as a front, leaving a cluster
of immune sites behind. The transition between survival and extinction
of the spreading agent is then described by a different type of critical
behavior, belonging to the universality class of dynamical
percolation (DyP)~\cite{Stauffer}. In fact, the resulting cluster of
immune sites has the same (or even identical) structure as an
isotropic percolation cluster. For this reason DyP can be used as a
dynamical procedure to generate isotropic percolation clusters.

As an immediate generalization one may consider an epidemic process in
which the strength of immunization can be
varied~\cite{CardyGrassberger85}. For example, the initial
susceptibility to infections $p_0$ may be locally set to a
different value $p$ when the first infection is encountered. 
In the following we shall denote this process as epidemic process with finite
immunization which we abbreviate for convenience as EPFI. The phase diagram of the EPFI in two
spatial dimensions was studied in~\cite{GCR97}. As shown
in Fig.~\ref{phasediag}, it comprises three different phases, including the GEP
($p=0$) and DP ($p=p_0$) as special cases.

The horizontal phase transition line in~Fig.~\ref{phasediag} can be
explained as follows. Obviously, when starting from a
fully occupied lattice, all sites become immediately immune so that the
infection rate is everywhere equal to $p$. Trivially, the
dynamics is then precisely that of a DP process controlled by $p$. This
gives rise to a DP transition at $p=p_c$, independent of $p_0$, where $p_c$
denotes the critical value of an ordinary DP process. However,
starting with a localized infected seed in a non-immune environment, the
situation is more subtle, though the horizontal line still
exists. For example, slightly above the critical line, i.e. for
$p=p_c+\epsilon$ with $\epsilon \ll 1$, the interior of a surviving
cluster is essentially dominated by an ordinary supercritical
DP process. Since such a DP process is characterized on large scales by a 
finite density of active sites, there will be a finite chance for the 
process to survive in the limit $t\rightarrow\infty$~\cite{GCR97}.

The situation below the horizontal line ($p<p_c$) is different.
In at least two or more dimensions this part of the phase diagram displays {\em two}
distinct phases, namely, an absorbing phase, where the
process stops after some time, and a phase of annular growth, where an
expanding front of high activity survives with finite
probability. Performing a field-theoretic renormalization group study
close to the upper critical dimension of DyP, $d_c^{_{\rm DyP}}=6$, and
computing the corresponding critical exponents, Janssen showed that
the critical behavior along the line connecting the GEP and the
DP point in~Fig.~\ref{phasediag} is that of DyP~\cite{Janssen85}. In
this sense immunization is a relevant perturbation, 
driving the system away from the DP point in
Fig.~\ref{phasediag} towards GEP. (We note that in the case of several
competing infections with immunization, one observes a crossover back to
DP~\cite{mut}. Another four-state generalization leads to in interesting
tricritical phenomenon belonging to a different universality class~\cite{TriCrit}).

More recently epidemic spreading with immunization appeared in a
different context. Studying systems with infinitely many absorbing
states such as the non-diffusive pair contact
process~\cite{Jensen93,JensenDickman93}, Mu{\~n}oz {\it et al}
conjectured that these models are described by the same type of field
theory as the EPFI~\cite{MGDL96,MGD98}. Roughly speaking, the frozen absorbing
configurations generated by the process provide a local memory of
activity in the past which effectively acts in the same way as
immunization (or weakening). Meanwhile this conjecture is widely
accepted, although some questions concerning the upper critical
dimension are still debated~\cite{Wijland02,Lubeck02}.

While the mentioned field-theoretic results explain the transition line
between the GEP and the DP point close to $d_c^{_{\rm DyP}}=6$, the critical
properties along the
horizontal line and in the vicinity of the DP point are
less well understood. Simulating the Langevin equation of the
EPFI at critical reinfection rate
(corresponding to the horizontal line in Fig.~\ref{phasediag}) in 1+1
dimensions starting with a localized seed, L{\'o}pez and Mu{\~n}oz
initially expected continuously varying exponents~\cite{LopezMunoz97},
but refined simulations and approximations suggest that there is no
power-law scaling. Instead, the activity was found to vary as a
stretched exponential in 1+1 dimensions~\cite{GCR97,JimenezHinrichsen03}.

In this paper we study the epidemic process with finite immunization,
focussing on the influence of immunization in the
vicinity of the DP point, especially in higher
dimensions $d>2$. In Sec.~\ref{MC-simulations} we describe Monte
Carlo simulations that are applied in the following
sections to obtain numerical results. In Sec.~\ref{pheno} we summarize the
most important phenomenological features of the process. In
particular, we argue that the clusters of
immune sites at the DP point are {\em compact} for $d \leq
3$ and {\em asymptotically} compact for $d=4$ (see below). Consequently, a
recently introduced scaling argument
that suggested a stretched exponential decay on the horizontal line
for $p_0\ll p_c$ in $d=1$~\cite{JimenezHinrichsen03} should apply in
any dimension $d \leq 3$ and maybe for $d=4$ as well, since it is based essentially on the compactness
of the cluster of immune sites. This means that we expect a stretched exponential 
decay instead of power-law behavior not only in one but also in two, three
and perhaps four spatial dimensions. 
Moreover, for $d<4$, primary infections and reinfections scale differently whereas they show the same scaling
behavior for $d\geq 4$. As a result, we find that the phase transition line
connecting the GEP and the DP point should terminate in the DP point 
with a {\em finite} slope for $d\geq 4$ (in contrast to a vanishing slope 
for $d<4$, as can be seen in Fig.~\ref{phasediag}). In
Sec.~\ref{fieldtheory} we discuss the Langevin equation for EPFI in
order to describe how the properties of the
process change between the upper critical dimension of DP, $d^{_{\rm
    DP}}_c=4$, and DyP, $d^{_{\rm DyP}}_c=6$. Numerically determined
phase diagrams for $d>2$ are presented in
Sec.~\ref{phase_diagrams}, which support the predictions of
Sec.~\ref{pheno}. Though we expect the behavior along
the horizontal phase transition line to be non-universal (at least for
$d\leq 4$), it is nevertheless possible to identify an exponent $\mu$ for
the temporal correlation length in the case of $p=p_c$ and $p_0=p_c
-\epsilon$ for $\epsilon \ll 1$. As discussed in
Sec.~\ref{temporal_correlation}, this exponent differs
from the critical exponent for the temporal correlation length of
DP. The main results are summarized in
Sec.~\ref{summary}. A heuristic argument for the value of $\mu$ that is
in accordance with the numerical data is presented in \ref{MU}. Numerical estimates
for critical parameters and critical exponents for DP and DyP are
given in \ref{DP} and \ref{DYP}, respectively.          

%
%
\section{Monte Carlo simulations}\label{MC-simulations}
%
%
We perform Monte Carlo (MC) simulations of the epidemic process with finite
immunization on a simple $d$-dimensional cubic lattice using the model of 
Ref.~\cite{GCR97} generalized to arbitrary dimensions. The process is defined
such that it reduces to directed {\em bond} percolation at the DP
point and to dynamical {\em bond} percolation at the GEP point. In particular,
an active lattice site $i$ at time $t$ may activate each of its $d$ nearest
neighbors, denoted as $j$, at the next discrete time step $t+1$. If neighbor $j$ has never been
active before, activity is transmitted from site $i$ to site $j$ with the
primary infection probability $p_0$. Contrarily, if site $j$ has been active
at least once in the past the infection probability is given by $p$, called
reinfection probability. Each site stays active only for one time step.  

The process is initialized with an active seed at the origin at
time $t=0$ in a non-immune environment. Each run is stopped either
when the process dies out or when it reaches a
preset maximum time. We average over many runs with different realizations of
randomness. The lattice is always chosen large enough so that the process
never reaches its boundary. Hence, finite size effects are eliminated. As
usual in such 'seed simulations' we measure the survival probability $P_{\rm s}(t)$
that the process survives at least up to time $t$, the number of active sites
averaged over all runs $N(t)$, and the mean square spreading from the origin
$R^2(t)$ averaged over surviving runs. At criticality these quantities obey
power-laws~\cite{GrassbergerTorre79,Janssen03} according to 
\begin{equation}
\label{critseed}
P_{s}(t)\sim t^{-\delta}\ \,,\ \ N(t)\sim t^\theta\ \ ,\ \ R^2(t)\sim t^{2/z}\ \,.
\end{equation}
This behavior is observed for critical DP as well as for critical DyP. Of
course, the exponents in Eq.~(\ref{critseed}) are generally different in 
both cases. In addition, we also measure the number of
primary infections $n_{\rm p}$ (see below).

From a technical point of view, our simulations for $d\geq 3$ are based on the
routine presented in~\cite{grassberger2003}. Lattice sites are labeled by
64-bit-long integers. We do not initialize storage for a whole $L^d$ lattice since for
high dimensions it is only possible to simulate small lateral lattice sizes
$L$ with this method. Instead we use lists to store the individual positions 
of active and immune sites. To perform an update
from $t$ to $t+1$ we go through all active sites at time $t$ and activate
their neighbors at time $t+1$ with probability $p_0$ or $p$, respectively. The
activated sites are stored in a list while the formerly active sites recover. 
In the case of a first infection, the site is added to the list of immune
sites. In order to efficiently check whether a site is immune or
not, and whether it has already been activated during the actual update step,
we use a hashing algorithm as described in~\cite{grassberger2003}. Contrarily, the
simulations for $d=1,2$ are carried out in the usual way where one
initializes storage for a whole lattice. However, in this case we apply
bit-coding, i.e., a single 64-bit-long integer stores the states of 64
lattice sites (active/inactive or immune/non-immune).    

%
%
\section{Phenomenological properties}
\label{pheno}
%
%
\begin{figure}
\centerline{\includegraphics[width=90mm]{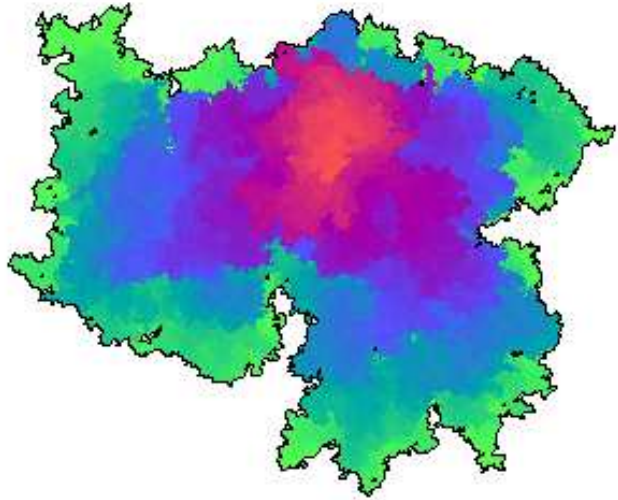}}
\caption{\label{cluster} Typical cluster of immune
sites generated by the EPFI in 2+1 dimensions
at the DP point after 2048 updates. The colors represent
linearly the time at which the sites were visited for the first
time. The black line marks the final boundary of the cluster. }
\end{figure}
Let us first consider the phenomenological properties of the process
near the DP point, where the effect of immunization is
small. Since the DP point itself is a critical DP process,
the question arises how the upper critical dimension of DP influences the
generated cluster of immune sites. As a conjecture we propose that the
generated cluster of immune sites at the DP point is
{\em compact} in $d < 4$ dimensions (in the sense that its fractal
dimension is $d$), while in higher dimensions $d>4$ it is not. 
At the DP point $p=p_0=p_c$, the influence of immunization vanishes so
that the cluster of immune sites is merely given by the sites visited
in the past by a critical DP process. In other words, when projecting
the past activity of a critical DP cluster onto space (looking 
through a critical DP cluster along the temporal axis) we expect its 
appearance to be compact in $d<4$. 
Obviously, in $d=1$ the cluster 
of visited sites is compact by definition so that the 
conjecture is correct. In dimensions $d=2,3$ the statement is
non-trivial. In fact, plotting the projection of a typical cluster 
in $d=2$ dimensions one obtains a compact object, as shown in 
Fig.~\ref{cluster}.

\begin{figure}
\centerline{\includegraphics[width=110mm]{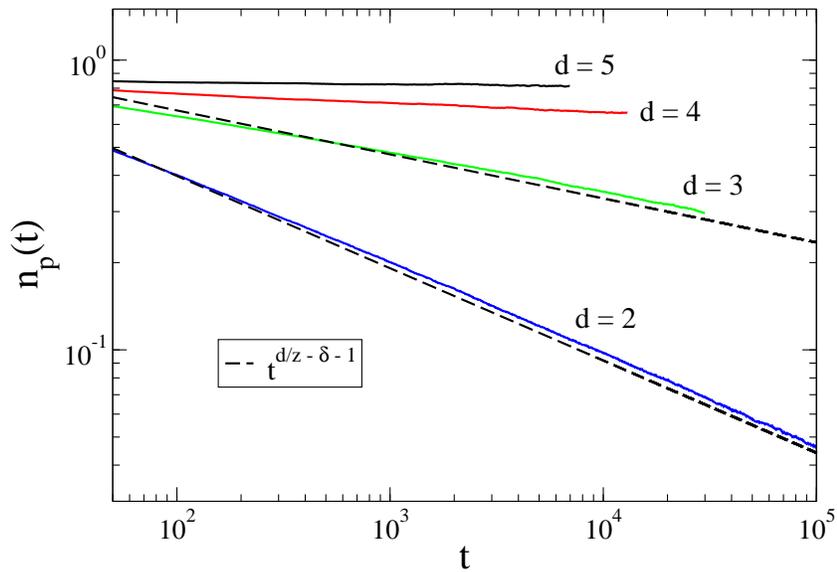}}
\caption{\label{f_infect}
Double logarithmic plot of the number of primary infections versus time at the
DP point for various dimensions. For comparison the power-law
behavior predicted by Eq.~(\ref{n_f}) is shown by the dashed lines for
$d=2,3$. For $d=4$ the mean-field prediction of Eq.~(\ref{n_f}) is $n_{\rm
  p}(t)=\text{const}$.}
\end{figure}

The above conjecture can be supported numerically as follows.
Assuming compactness, the number of immune sites in a surviving run
should grow linearly with the volume $\xi_\perp^d$, where $\xi_\perp
\sim t^{1/z}$ is the spatial correlation length and
$z=\nu_\parallel/\nu_\perp$ is the dynamical exponent of DP (see,
e.g.,~\cite{Hinrichsen}). Averaging over all runs, the volume has to be 
multiplied with the survival probability $P_s(t) \sim t^{-\delta}$, where
$\delta=\beta/\nu_\parallel$, hence the average number of immune sites
$I(t)$ increases as
\begin{equation}\label{I}
I(t) \sim t^{d/z-\delta}\,.
\end{equation}
This implies that the number of {\em primary} infections $n_{\rm p}$ scales as
the derivative of $I(t)$, i.e.,
\begin{equation}\label{n_f}
n_{\rm p}(t) \sim t^{d/z-\delta -1}\,.
\end{equation}
As shown in Fig.~\ref{f_infect} by numerical simulations, in $d=2,3$
this quantity scales indeed according to Eq.~(\ref{n_f}), supporting
that the generated clusters of immune sites are compact.

For $d=4$ the data still exhibits a slight curvature which is presumably due
to logarithmic corrections~\cite{JanssenLog,Lubeck04} that affect the mean-field behavior ($n_{\rm
  p}=\text{const}$) at the upper critical dimension of DP, $d_c^{_{\rm
    DP}}=4$. As usual, also for $d>d_c^{_{\rm DP}}$
we observe mean-field behavior and thus we expect 
the resulting cluster of immune sites to be non-compact.

It is also possible to support the conjecture by a scaling argument.
Obviously $I(t)$ cannot be larger than the integrated past activity $\int_0^t dt'N(t') \sim
t^{1+\theta}$~\cite{GrassbergerTorre79}, leading to the inequality
\begin{equation}\label{in_equ}
d/z-\delta \leq \theta +1\ .
\end{equation}
as a necessary condition for compactness. In $d<4$ dimensions the initial-slip 
exponent $\theta$ is given by the hyperscaling relation~\cite{GrassbergerTorre79,munoz}
\begin{equation}
\label{Hyperscaling}
\theta = d/z-2\delta
\end{equation}
so that the inequality reduces to $\delta \leq 1$, which is indeed satisfied 
in $d<4$ (see Tab.~\ref{TableDP}). At the upper critical dimension of DP, $d_c^{_{\rm DP}}=4$, 
taking the mean-field exponents $\theta=0$, $z=2$, and $\delta=1$, the inequality (\ref{in_equ}) 
is sharply satisfied while it is violated above four dimensions. Hence, we can conclude that in $d>4$ 
dimensions the generated cluster is no longer compact.

To deal with the question whether clusters of immune sites in systems {\em at} 
the upper critical dimension of DP, $d_c^{_{\rm DP}}=4$, are still
compact, it is worthwhile to compare the situation with a simple random walk whose upper
critical dimension is  $d_c^{_{\rm RW}}=2$. The random walk is {\em
  recurrent} in $d=1,2$ while it is {\em transient} for $d\geq
3$~\cite{RW}. This means that the probability $F(t)$ to visit a given
site at least once until time $t$ tends to $1$ in the limit $t\to\infty$ for $d=1,2$ while it tends to a
constant $c<1$ for $d\geq 3$. Therefore, the region visited by the
random walker is compact in $d\leq 2$ whereas it is not in $d\geq
3$. However, for $d=2$ the probability $1-F(t)$ that the site has not
yet been visited decreases asymptotically as $1-F(t)\sim(\ln t)^{-1}$ in contrast to algebraic
behavior in $d=1$, where $1-F(t)\sim t^{-1/2}$. This implies that in $d=2$ compactness
is reached only slowly and we refer to this as {\em asymptotically}
compact. Therefore, in $d=2$ and finite time the cluster of visited
sites is highly non-compact at its boundaries and differs significantly
from the one shown in Fig.~\ref{cluster}. Using this analogy, it is
near at hand to speculate that the same happens in a
critical DP process. Thus, we expect the cluster of visited sites to
be compact at the upper critical dimension $d_c^{_{\rm DP}}=4$ as
well. However, in this case compactness should be reached considerably slower than
for $d<4$, i.e., we expect the clusters to be {\em asymptotically} compact.
 
The observed compactness in low dimensions
leads to an important consequence regarding the
critical behavior along the horizontal line in~Fig.~\ref{phasediag},
as it allows the results of~\cite{JimenezHinrichsen03} obtained in $d=1$
to be generalized to $2$ and $3$ dimensions. In~\cite{JimenezHinrichsen03} the
expansion of an immune region in one spatial dimension was studied in
the limit of a very small probability of first infections $p_0 \rightarrow 0$. 
Using a quasi-static approximation it was shown that the survival
probability does {\em not} obey a power law, instead it decays as a
stretched exponential
\begin{equation}
\label{SurvProb}
P_s(t) \propto \exp\left( -A p_0^{-\alpha} t^{1-\alpha} \right),
\end{equation}
giving rise to a {\em finite} survival time $T\sim
p_0^{\alpha/(1-\alpha)}$. Here
$\alpha=\nu_\parallel/(\nu_\perp+\beta_s)$ with $\beta_s$ being the
order parameter exponent next to a planar absorbing
surface~\cite{FHL01}. The main assumption made in this approximation
is the compactness of the immune domain, wherefore the analysis had been
restricted to $d=1$. Because of the observed compactness, the
same arguments can be applied in $2$ and $3$ dimensions so that 
the formula Eq.~(\ref{SurvProb}) should be valid in these cases as well. 
Inserting the known values for $\beta_s$~\cite{FHL01} one obtains 
$\alpha\simeq0.947$ in 1$d$ and $\alpha\simeq 0.72$ in 2$d$ (so far
there are no estimates of $\beta_s$ in 3$d$). Whether Eq.~(\ref{SurvProb})
is still valid at the upper critical dimension of DP, $d_c^{_{\rm DP}}=4$,
where $\alpha$ would be $1/2$, is not yet clear since in this case we expect
the immune region to be only asymptotically compact.
The numerical verification of the stretched exponential behavior for
$d=1$ and $p_0 \ll p_c$ in~\cite{JimenezHinrichsen03} required
a special enrichment method since for small $p_0$ most simulation runs
terminate after a very short time. In this paper we perform ordinary
Monte Carlo simulations and therefore we do not address the numerical
analysis of the horizontal line for $p_0 \ll p_c$ in $d>1$. 

Another important consequence concerns the ratio of first infections and
reinfections. In $d<4$ dimensions the average number of first infections $n_p(t)$
(controlled by the parameter $p_0$) decreases according to Eq.~(\ref{n_f})
while the average number of reinfections $N(t)$ 
(controlled by the parameter $p$) increases as $t^\theta$ with $\theta>0$.
Thus, after sufficiently long time the parameter $p$ controlling reinfections will
have a much larger influence than the parameter $p_0$. As a consequence
the curved phase transition line, which may be thought of as describing a situation
where $p$ and $p_0$ balance each other, terminates in the DP point horizontally,
i.e. with zero slope. In $d\geq 4$ dimensions, primary infections and reinfections 
show the same scaling behavior (for $d=4$ probably affected by logarithmic
corrections), and therefore we expect the transition line to terminate with
a non-vanishing slope. We will come back to this question in Sec.~\ref{temporal_correlation}.
%
%
\section{Scaling properties}
\label{fieldtheory}
%
%
The previous phenomenological arguments suggested that the behavior of the model
in $d<4$ dimensions differs significantly from the behavior above $4$ dimensions.
This difference becomes also obvious when studying the corresponding Langevin
equations. 

According to Refs.~\cite{Janssen85,JimenezHinrichsen03} the Langevin equation for the
epidemic process with finite immunization reads
\begin{eqnarray}
\label{Langevin}
\frac{\partial}{\partial t} \rho(\xvec,t) &=& a \rho(\xvec,t) - b
\rho^2(\xvec,t) + D \nabla \rho(\xvec,t) + \xi(\xvec,t) \\ &&+ \lambda
\, \rho(\xvec,t) \, \exp\left(-w \int_0^t d\tau \, \rho(\xvec,\tau)
\right)\,, \nonumber
\end{eqnarray}
where $\xi(\xvec,t)$ represents a density-dependent Gaussian noise
with the correlations
\begin{equation}
\langle \xi(\xvec,t) \xi(\xvec',t') \rangle = \Gamma\, \rho(\xvec,t)
\, \delta^d (\xvec-\xvec') \, \delta(t-t').
\end{equation}
It consists of the usual Langevin equation for DP plus an exponential
term describing the effect of immunization. Here the exponential
function can be thought of as a switch: Initially, the integral is
zero and hence the coefficient of the linear contributions in
$\rho(\xvec,t)$ is $a+\lambda$, representing as bare parameters
the reduced rate for first infections $p_0-p_c$. 
However, when the integrated past activity exceeds $1/w$,
the exponential function decreases rapidly so that the additional term
is essentially switched off. Roughly speaking, in a continuous
description the parameter $w$ is needed in order to specify a
threshold telling us how much activity has to be accumulated
at a given site in order to declare it as immune. Once the
exponential term is switched off, the linear term is controlled by the
coefficient $a$ which represents the reduced reinfection rate
$p-p_c$. In most lattice models, sites become immune after a single
infection, hence the parameter $w$ is of the order $1$ while $\lambda
\sim p_0-p$ controls the strength of immunization.

Rescaling the DP Langevin equation by
\begin{equation}
\xvec \to b \xvec\, \quad t \to b^z t \, \quad \rho \to b^{-\chi} \rho
\end{equation}
with a scaling parameter $b$ and the exponents
$z=\nu_\parallel/\nu_\perp$ and $\chi=\beta/\nu_\perp$ one immediately
recognizes that simple scaling invariance at the upper critical
dimension $d_c^{_{\rm DP}}=4$ can only be established if $z=\chi=2$
and $a=0$, the latter representing the critical point at the mean
field level. Regarding the exponential term, scaling invariance
requires the argument of the exponential function and the
exponential function itself to be dimensionless, hence the coefficient
$\lambda$ and $w$ have to be rescaled as
\begin{eqnarray}
\lambda &\to & b^{-y_\lambda} \lambda \\ w &\to & b^{-y_w} w \nonumber\,,
\end{eqnarray}
where $y_\lambda=2$ and $y_w=0$. More generally, it can be shown by a field-theoretic
calculation~\cite{JimenezHinrichsen03} that in $d\leq 4$ dimensions
the two exponents are given by
\begin{equation}
y_\lambda = \frac{1}{\nu_\perp} \,, \qquad y_w =
\frac{\nu_\parallel-\beta}{\nu_\perp}\,.
\end{equation}
As $y_w$ and $y_\lambda$ are positive, the exponential term for immunization is relevant
in $d \leq 4$ dimensions.  Expanding the exponential function 
as a Taylor series, the resulting terms would be {\rm equally} relevant 
in $d=4$ and {\em increasingly relevant} in $d<4$ dimensions. Therefore,
in $d \leq 4$ dimensions a Taylor expansion of the exponential function in
Eq.~(\ref{Langevin}) is meaningless in the renormalization group sense,
instead it has to be kept in as a whole. This circumstance is probably
responsible for the observed non-universality along the horizontal
line in Fig.~\ref{phasediag}.

In $d>4$ dimensions, however, the situation is different. Here,
power counting at the upper critical dimension
of DyP, $d_c^{_{\rm DyP}}=6$, yields $y_w<0$, meaning that the
relevancy of the terms
in the Taylor expansion decreases. In this case it is
legitimate to carry out the Taylor expansion, keeping only the most
relevant contribution. As the zeroth order can always be absorbed in
a redefinition of~$a$, the most relevant contribution is the
first-order term $-\lambda w \rho(\xvec,t) \int_0^t d\tau
\rho(\xvec,\tau)$. In addition we may drop the quadratic contribution
$-b\rho^2(\xvec,t)$, which is irrelevant at $d_c^{_{\rm
DyP}}=6$, leading to the Langevin equation
\begin{eqnarray}
\label{Langevin2}
\frac{\partial}{\partial t} \rho(\xvec,t) &=& (a+\lambda)
\rho(\xvec,t) + D \nabla \rho(\xvec,t) + \xi(\xvec,t)\\ &&- \lambda w\,
\rho(\xvec,t) \int_0^t d\tau \, \rho(\xvec,\tau) \,.\nonumber
\end{eqnarray}
corresponding to the field theory studied in~\cite{CardyGrassberger85,Janssen85}. In
contrast to Eq.~(\ref{Langevin}) the influence of immunization is
effectively described by a single parameter, namely, by the product
$\lambda w$.  Assuming $y_w$ to be negative for any $d>4$ this
scenario is expected to hold even in presence of fluctuation effects.
Moreover, in contrast to Eq.~(\ref{Langevin}) the parameter now 
appears in the prefactor of the linear term, shifting the parameter $a$.
This suggests, in accordance with the preceding section, that at
least in $4<d\leq 6$ dimensions the reduced first infection and reinfection
probability (here corresponding to $\lambda$ and $a$) scale identically
and that the phase transition line, along which the influence of $a$ and
$\lambda$ is balanced, terminates with a nonzero slope in the DP point.\\

To summarize we arrive at the following picture:

\begin{itemize}
\item
In $d \leq 4$ dimensions the EPFI is described by
Eq.~(\ref{Langevin}), in which an expansion of the exponential
function is not allowed. The exponential function is conjectured to
produce non-universal features such as a stretched exponential decay
of the survival probability, as observed in $d=1$
dimensions~\cite{JimenezHinrichsen03}.

\item
In $4 < d \leq 6$ the process is described by
Eq.~(\ref{Langevin2}). Fluctuation effects are still present and  the
corresponding field theory is well-defined and
renormalizable~\cite{Janssen85}.

\item
In $d > 6$ the mean field approximation becomes valid. The system is 
driven to a trivial Gaussian fixed point, which is the same for GEP and DP.
\end{itemize}
%
%
\section{Phase Diagrams}
\label{phase_diagrams}
%
%
\begin{figure}
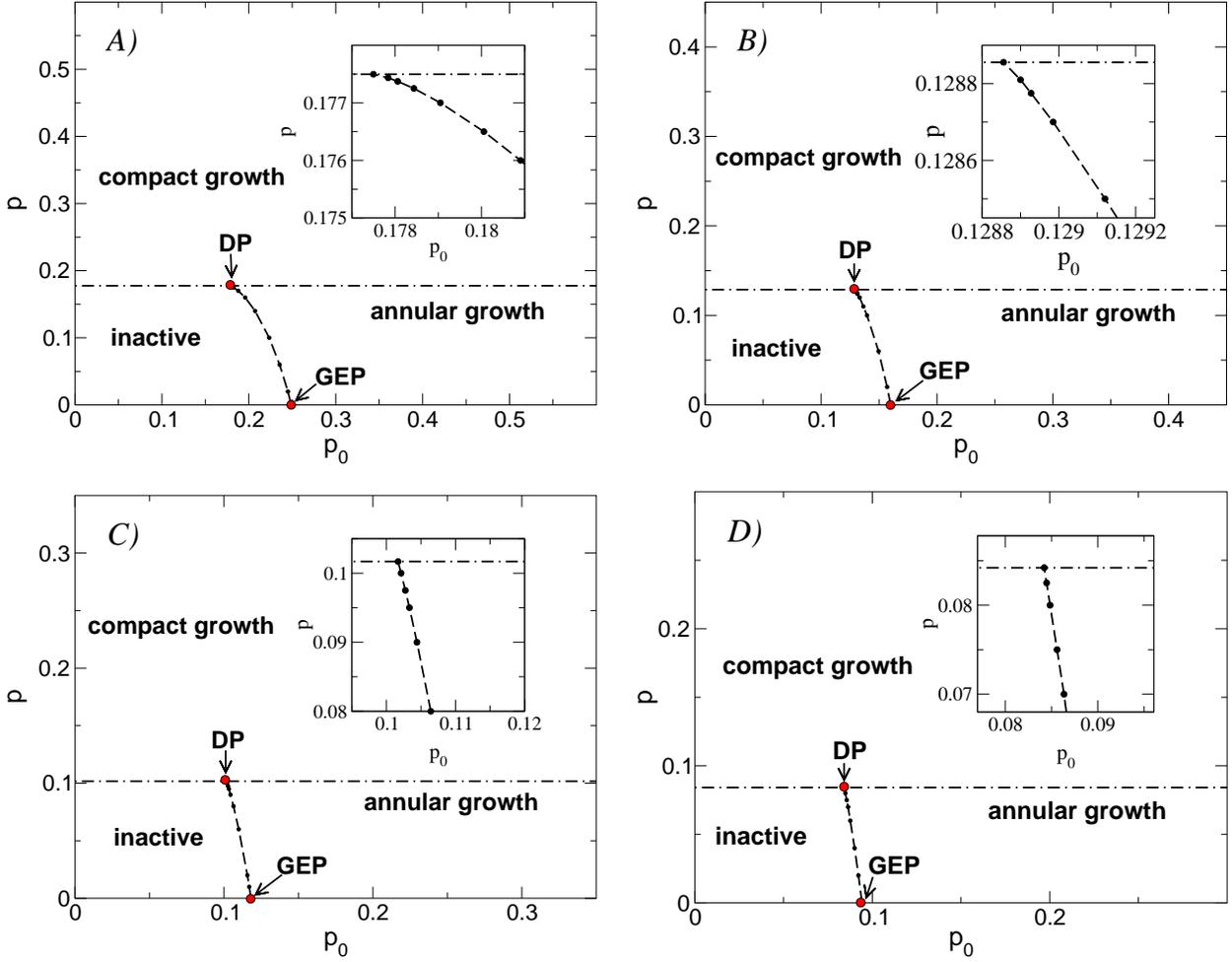
 
\includegraphics[width=80mm]{phase3d.eps}\hglue5mm
\includegraphics[width=80mm]{phase4d.eps}\vspace{3mm}\\
\includegraphics[width=80mm]{phase5d.eps}\hglue5mm
\includegraphics[width=80mm]{phase6d.eps}
\caption{
\label{phasediagrams}
Phase diagrams for the EPFI for $d=3,4,5,6$, $A)$--$D)$, determined from MC
simulations. The insets show the vicinity of the DP point. Circles
mark the numerically determined critical points.  
}
\end{figure} 
Fig.~\ref{phasediagrams} presents phase diagrams of the epidemic process
with finite immunization for spatial dimensions $d=3,4,5,6$ which were
obtained obtained from MC simulations (see
Sec.~\ref{MC-simulations}). The horizontal lines obey $p=p_c$ with the values
of $p_c$ given in Tab.~\ref{TableDP}. The lines connecting the
DP and the GEP point were determined as follows.
First we performed spreading simulations at the critical point of the GEP, i.e.,
for $p=0$, using the critical values of bond percolation given
in~\cite{grassberger2003}. Thereby, we estimated  various critical
exponents for DyP which are presented in Tab.~\ref{TableGEP}. 
In order to locate the phase transition line, we then used the fact
that the critical behavior
along this line is that of dynamical percolation. We determined various
critical points along the lines by keeping $p$ fixed and varying $p_0$ until
the quantities $P_s(t)$, $N(t)$ and $R^2(t)$, see Eq.~(\ref{critseed}), displayed the
expected slope corresponding to DyP in a log-log plot. 

In accordance with the predictions of Sec.~\ref{pheno} the data indeed
suggests that the curved phase transition line terminates at the DP point
with vanishing slope for $d=3$ and with finite slope for $d\geq 4$. For $d=4$
this behavior is not so clear cut and we believe that this is caused by
logarithmic corrections at the DP point.   

Assuming that the curved phase transition line behaves in the
vicinity of the DP point as $p_{0,c}(p)-p_c\sim
(p_c-p)^\gamma$, our data leads to $\gamma = 0.47(7)$ for $d=2$ and
to $\gamma = 0.74(8)$ for $d=3$. Since the error bars are rather large, our
data does only allow a rough estimate of $\gamma$ and we did not
further attempt to relate its value to other critical exponents.
%
%
\section{Temporal correlation length in the vicinity of the DP point}
\label{temporal_correlation}
%
%
%
%
\begin{figure}[t]
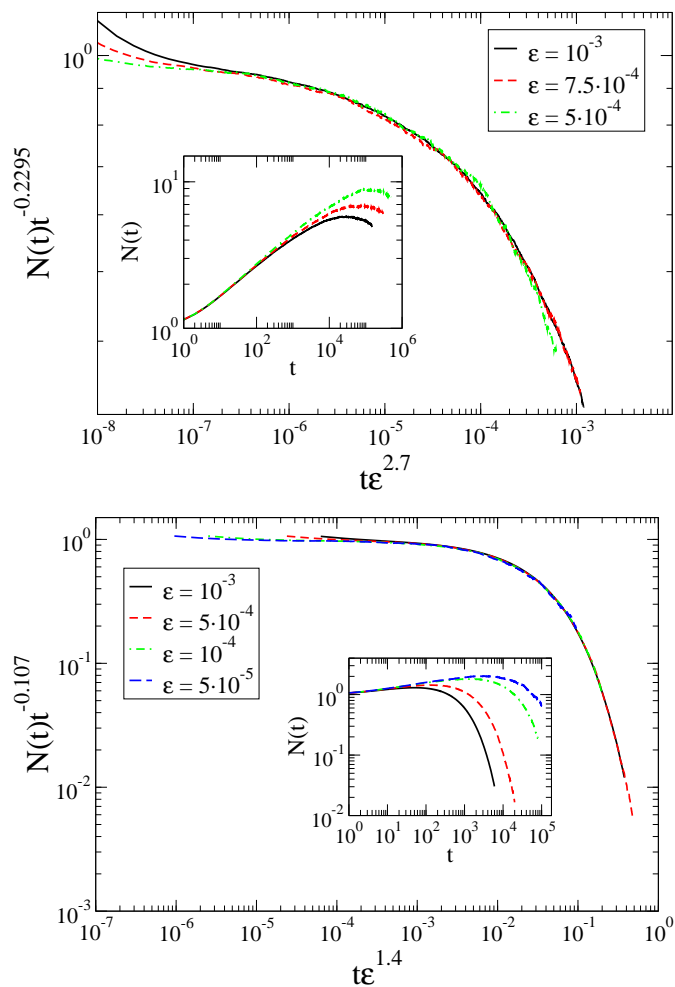
 
\centerline{\includegraphics[width=88mm]{left2d.eps}}\vspace{3mm}
\centerline{\includegraphics[width=88mm]{left3d.eps}}
\caption{
\label{left}
Data collapses for $p=p_c$ and $p_0=p_c-\epsilon$ for $d=2$ (top) and
$d=3$ (bottom). $N(t)$ is scaled with $t^\theta$ with the values of
$\theta$ from Tab.~\ref{TableDP}. The time is scaled with $\epsilon ^\mu$ such
that the best data collapses were obtained. The insets show the original data.
}
\end{figure} 
Off criticality but in the vicinity of the DP point, i.e., $p=p_c\pm\tilde{\epsilon}$,
$p_0=p_c\pm\epsilon$, one initially observes the critical behavior of
DP until the process eventually crosses over to a different type
of behavior. This crossover takes place at a certain typical
time scale $\xi _\parallel$. If $\tilde{\epsilon}\neq 0$ 
(moving vertically away from the DP point) one expects that asymptotically $\xi
_\parallel\sim\tilde{\epsilon}^{-\nu _\parallel}$ with the critical
DP exponent $\nu _\parallel$ for the temporal correlation
length. The reason is the following. Near the DP point, for $d < 4$,
the process is dominated by reinfections that perform an ordinary
DP process with parameter $p$ on a compact region (see
Sec.~\ref{pheno}). It is only on the edge of the visited region where
the first infection probability $p_0$ plays a role. Hence, the time
scale associated with $\tilde{\epsilon}$ is expected to be smaller than
that associated with a change $\epsilon$ of $p_0$. For $d\geq 4$, primary
infections and reinfection show equal scaling behavior and therefore one
expects both times to scale in the same way. 

We are left with the question how the temporal correlation length
behaves for $p=p_c$  and $p_0=p_c\pm\epsilon$ (moving horizontally away from the DP point)
in $d<4$. Although we expect non-universal behavior on the horizontal line it may 
nevertheless be possible to identify an exponent $\mu$ for the temporal correlation 
length
\begin{equation}
\xi_\parallel\sim \epsilon^{-\mu}
\end{equation}
in the vicinity of the DP point, $\epsilon\ll 1$. In fact, as shown in
Fig.~\ref{left}, plotting $N(t)t^{-\theta}$ versus $t/\xi_\parallel$ 
we can produce a data collapse by tuning $\mu$. With similar simulations 
in $d=4,5$ (not shown here) we get the estimates
\begin{equation}
\mu = \left\{
\begin{array}{cc}
2.7(7) & \text{ in } d=2 \\
1.4(2) & \text{ in } d=3 \\
1.1(13) & \text{ in } d=4 \\
1.0(1) & \text{ in } d=5
\end{array}
\right.
\end{equation}
Obviously, in $d<4$ the exponent $\mu$ differs from $\nu_\parallel$ ($\nu _\parallel=1.295$ for
$d=2$~\cite{voigt_ziff} and $\nu _\parallel=1.105$~\cite{Jensen92} for
$d=3$). Because of logarithmic corrections, it is likely that $\mu=1$  in $d=4$ as well.
With decreasing dimension the value of $\mu$ and therewith the simulation
time increases rapidly. For this reason the estimate for $d=2$ is less 
precise, while simulations in $d=1$ turned out to be unreliable. However, for $d=3$ 
the data collapse is fairly good. In App.~\ref{MU} we present a heuristic argument
for the value of $\mu$ in accordance with the numerical data.
%
%
\section{Summary}
\label{summary}
%
%
We studied a model for spreading with finite immunization
which is controlled by probabilities for first infections ($p_0$) and
reinfections ($p$). We focussed in particular on the critical behavior close
to the directed percolation point, especially in high
dimensions $d>2$. We argued that we expect the domains of immune sites to
be compact for $d\leq 4$, however, with an approach to compactness
that is (logarithmically) slow at the upper critical dimension of DP, $d^{_{\rm
    DP}}_c=4$. Therefore, we denoted the visited region as asymptotically
compact in $d=4$. The compactness of the
immune region was supported by MC-simulations. We pointed out that this
compactness implies that a recently introduced scaling argument, suggesting a
stretched exponential decay of the survival probability for $p=p_c$, $p_0\ll
p_c$ in one spatial dimension, should apply in any dimension $d \leq 3$ and
maybe in $d=4$ as well. Furthermore, we showed that for $d<4$ the number of
first infections (averaged over all runs) decreases whereas the number of
reinfections increases. Contrarily, for $d\geq 4$ both quantities scale
equally. From that we derived the result that the phase transition line
connecting the GEP and the DP point terminates in the DP point 
with a {\em finite} slope for $d\geq 4$ and with a vanishing slope 
for $d<4$, which was supported by numerically determined phase diagrams.
We also discussed the Langevin equation for the process, in order to
study how the properties of the process change depending on the spatial
dimension. Investigating the behavior for $p=p_c$ and $p_0=p_c-\epsilon$, $\epsilon\ll 1$
we were able to identify an exponent $\mu$ for the temporal correlation length which is
different from $\nu _\parallel$ for DP. 


\vglue3mm
\noindent {\bf Acknowledgments:}\\
Parts of the numerical simulations were carried out on the 128-node
Alpha Linux Cluster Engine ALICE at the University of Wuppertal. We
thank N. Eicker, T. Lippert, and B. Orth for technical
support. Moreover, we are grateful to two referees for
bringing references~\cite{reweight,ziff,Grassberger92} to our
attention.

\vglue10mm
\appendix

\section{Speculation about the value of $\mu$}\label{MU}
In the following we present a heuristic argument for the value of
$\mu$ in $d\leq 4$ dimensions that is in accordance with the 
numerical results presented above. Consider first a
subcritical DP process governed by the parameter $p=p_c-\epsilon$, $\epsilon \ll
1$. Initially the system behaves as if it was critical until the hostile
conditions eventually lead to extinction. This happens on a
time scale $\xi _\parallel\sim\epsilon ^{-\nu_\parallel}$. During the active
time the process produces active sites according to $N(t)\sim t^\theta$, see
Eq.~(\ref{critseed}). Therefore, a subcritical DP process activates on the
whole $M$ sites before it dies out, where $M$ scales as $M\sim\epsilon
^{-\nu_\parallel (\theta +1)}$. Consider now the EPFI with $p=p_c$ and
$p_0=p_c-\epsilon$ which dies out on a time scale $\xi _\parallel\sim\epsilon
^{-\mu}$. In this case (for $d\leq 4$) one has basically a critical DP process
in the compact interior region of the immune cluster and a subcritical
process on its edge. The integrated activity on the edge scales as $I(t) \sim
t^{d/z-\delta}$, see Eq.~(\ref{I}). If one assumes that the critical
reinfections do not introduce a time scale and that the process on the edge is
subject to the same bound as subcritical DP concerning the maximal activity,
one arrives at $I\sim M$ leading to
\begin{equation}\label{mu}
\mu = \frac{z\nu_\parallel (\theta +1)}{d-\delta z}\ .
\end{equation}       
With the critical exponents of DP given in Sec.~\ref{temporal_correlation} and in Tab.~\ref{TableDP}
Eq.~(\ref{mu}) results in $\mu \approx 2.335$ for $d=2$, $\mu = 1.437$ for
$d=3$ and $\mu = 1$ for $d=4$ which is compatible with the numerical
results. For $d=1$, with Tab.~\ref{TableDP} and $\nu _\parallel =
1.7338$~\cite{Jensen99} one obtains $\mu \approx 4.814$. However,
this argument needs to be substantiated and therefore 
Eq.~(\ref{mu}) has to be taken with care.

\section{DP in high dimensions}\label{DP}
In this Appendix we report some numerical results on directed bond
percolation in high dimensions. The values are partly more precise than previously
reported estimates. Similar results for directed site percolation in
$d=3,4$, and $5$ dimensions  were recently reported
in~\cite{Lubeck02,Lubeck04}.

\begin{table}\small \center
\begin{tabular}{|c||c|c|c|c|c|}
\hline $\quad d\quad $ & $p_c$ & $\delta$ & $\theta$ & $z$ & Ref. \\
\hline $1$ & 0.644700185(5) & 0.159464(6) & 0.313686(8) & 1.580745(10)
& \cite{Jensen99}\\ 
$2$ & 0.287338(3)$^\dagger$ & 0.4505(10) & 0.2295(10) & 1.766(2) & \cite{voigt_ziff} \\ 
$3$ & 0.1774970(5) &
0.737(5) & 0.107(5) & 1.897(5) & present work \\ 
$4$ & 0.1288557(5) &
$1^*$ & $0^*$ & $2^*$ & present work \\ 
$5$ & 0.1016796(5) & 1 & 0 & 2
& present work \\ 
$6$ & 0.0841997(14) & 1 & 0 & 2 & present work \\ $7$ &
0.07195(5) & 1 & 0 & 2 & present work \\ \hline
\end{tabular}
\caption{\label{TableDP} Numerical results for directed bond
percolation on a simple $d$-dimensional cubic lattice. For $d=4$, $^*$
denotes that the mean-field behavior is subjected to logarithmic
corrections. $z=\nu _\parallel / \nu _\perp$. $\dagger$ Taken from~\cite{Grassberger96}.}
\end{table}

Table~\ref{TableDP} shows the results for directed bond percolation
ranging  from $1$ to $7$ spatial dimensions. Our estimates for $d=3$
differ slightly from that presented previously
by Dickman~\cite{reweight}, namely, $\delta = 0.7263(11)$, $\theta = 0.110(1)$
and $z=1.919(4)$. Though the error bars in~\cite{reweight} are smaller
than in our case, our data is incompatible with the values of $\delta$
and $z$ in~\cite{reweight}. We do not conclude with respect to this issue but
note that further numerical effort in this context is desirable.   
The value $\theta=0.107(5)$ is smaller than Dickmans result $\theta=0.110(1)$ but
larger than the field-theoretic estimate
$\theta=0.098$~\cite{Janssen81} based on a two-loop calculation.

Finally we note that our estimates in $3d$ are compatible with the DP
hyperscaling relation~\cite{GrassbergerTorre79,munoz} where, inserting our
estimates for $z$ and $\delta$, we obtain
\begin{equation}
\theta = d/z - 2\delta = 0.107(10)
\end{equation}
In $d>4$, however, the hyperscaling relation is violated.

\section{Dynamical percolation in high dimensions}\label{DYP}
Here we present our numerical estimates for the values of critical thresholds and
exponents for dynamical bond percolation on a $d$-dimensional cubic lattice. Note
that, as far as we know, no numerical values for critical exponents for DyP in
$d=4,5$ have been published before. The values presented for
$d=3$ are compatible with precise estimates published previously, i.e.,
$p_c =0.2488126(5)$~\cite{ziff}, $\delta =0.345(4)$,
$\theta =0.494(6)$ and $1/z=0.728(2)$ where the latter values are
from~\cite{Grassberger92}. 
\begin{table}\small \center
\begin{tabular}{|c||c|c|c|c|c|}
\hline $\quad d\quad $ & $p_c$ & $\delta$ & $\theta$ & $z$ & Ref. \\\hline
$2$ & 0.5$^\dagger$ & 0.092 & 0.586 & 1.1295 & ~\cite{munoz2}$^{\ddagger}$ \\
$3$ & 0.2488125(25) & 0.346(6) & 0.488(7) & 1.375(5) & present work \\ 
$4$ & 0.1601310(10) & 0.595(8) & 0.30(1) & 1.605(9) & present work \\ 
$5$ & 0.1181718(3) & 0.806(12) & 0.134(10) & 1.815(10)& present work \\ 
$6$ & 0.0942019(6) & $1^*$ & $0^*$ & $2^*$ & \cite{grassberger2003} \\ \hline
\end{tabular}
\caption{\label{TableGEP} Numerical results for dynamical bond percolation on a
  simple $d$-dimensional cubic lattice. For $d=6$, $^*$ denotes that the
  mean-field behavior is subjected to logarithmic corrections. $z=\nu
  _\parallel / \nu _\perp$. $^\dagger$ Taken
  from~\cite{Stauffer}. $^\ddagger$ Uncertainties are in the last
  digit. For $d=1$ the transition is shifted to $p_c=1$.}
\end{table}

The value of $\theta=0.488(7)$ is incompatible with the value of $\theta = 0.536$
reported in~\cite{munoz2} based on previous estimates
from~\cite{bunde}. However, in this case $\theta$ was not directly
measured but obtained via scaling relations, which presumably is the reason
for the considerable difference between the two estimates.

The estimates given in Tab.~\ref{TableGEP} are compatible with the
hyperscaling relation for DyP~\cite{munoz,Grassberger92}. For $d=3,4,5$ we obtain
\begin{equation} 
\theta = d/z-2\delta -1 = \left\{
\begin{array}{cc}
0.49(2) & \text{for } d=3\ ,\\
0.30(3) & \text{for } d=4\ ,\\
0.14(4) & \text{for } d=5\ .
\end{array}
\right.
\end{equation}

\newpage

\noindent{\bf References}\\


\begin{thebibliography}{99}                               

\bibitem{MarroDickman}  
Marro J and Dickman R, 
\textit{Nonequilibrium phase transitions in lattice models}, 
Cambridge University Press, Cambridge (1999).

\bibitem{Hinrichsen}  
Hinrichsen H, 
\textit{Non-equilibrium critical phenomena and phase transitions into absorbing states},
2000 Adv. Phys. {\bf 49}, 815 [cond-mat/0001070].

\bibitem{OdorReview} 
{\'O}dor G, 
\textit{Universality classes in nonequilibrium lattice systems},
submitted [cond-mat/0205644v7].

\bibitem{Brazil} 
Hinrichsen H, 
\textit{On possible experimental realizations of Directed Percolation},
2000 Braz. J. Phys. {\bf 30}, 69 [cond-mat/9910284].

\bibitem{Mollison77} 
Mollison D, 
\textit{Spatial contact models for ecological and epidemic spread},
1977 J. R. Stat. Soc. B {\bf 39}, 283.

\bibitem{mut} 
Dammer S M and Hinrichsen H, 
\textit{Epidemic spreading with immunization and mutations},
2003 Phys. Rev. E {\bf 68}, 016114 [cond-mat/0303467].

\bibitem{ZGB86} 
Ziff R M, Gulari E, and Barshad Y,
\textit{Kinetic phase-transitions in an irreversible surface-reaction model},
1986 Phys. Rev. Lett. {\bf 56}, 2553.

\bibitem{HJRD99} 
Hinrichsen H, Jim\'enez-Dalmaroni A, Rozov Y, and Domany E,
\textit{Flowing sand - a physical realization of Directed Percolation},
1999 Phys. Rev. Lett. {\bf 83}, 4999 [cond-mat/9908103].

\bibitem{HJRD99b} 
Hinrichsen H, Jim\'enez-Dalmaroni A, Rozov Y, and Domany E,
\textit{Flowing sand - a possible physical realization of Directed Percolation},
2000 J. Stat. Phys. {\bf 98}, 1149 [cond-mat/9909376].

\bibitem{Kinzel85}  
Kinzel W, 
\textit{Phase transitions of cellular automata},
1985 Z. Phys. B {\bf 58}, 229.

\bibitem{Reggeon1}  
Moshe M, 
\textit{Recent developments in Reggeon field theory},
1978 Phys. Rep. C {\bf 37}, 255.

\bibitem{Reggeon2}  
Grassberger P and Sundermeyer K, 
\textit{Reggeon field theory and Markov processes},
1978 Phys. Lett. B {\bf 77}, 220.

\bibitem{Reggeon3}  
Cardy J L and Sugar R L, 
\textit{Directed percolation and Reggeon field theory},
1980 J. Phys. A {\bf 13}, L423.

\bibitem{Grassberger83}  
Grassberger P, 
\textit{On the critical behavior of the general epidemic process and dynamical percolation},
1983 Math. Biosci. {\bf 63}, 157.

\bibitem{Stauffer}  
Stauffer D and Aharony A,
\textit{Introduction to Percolation Theory}, 
2nd ed., Taylor \& Francis, London (1992).

\bibitem{CardyGrassberger85} 
Cardy J L  and Grassberger P,
\textit{Epidemic models and percolation},
1985 J. Phys. A {\bf 18}, L267.

\bibitem{GCR97}  
Grassberger P, Chat{\'e} H, and Rousseau G,
\textit{Spreading in media with long-time memory},
1997 Phys. Rev. E {\bf 55}, 2488.

\bibitem{Janssen85}  
Janssen H K,
\textit{Renormalized field theory of dynamical percolation},
1985 Z. Phys. B {\bf 58}, 311.

\bibitem{TriCrit}
Janssen H K, M{\"u}ller M, and Stenull O,
\textit{A Generalized Epidemic Process and Tricritical Dynamic Percolation},
submitted [cond-mat/0404167].

\bibitem{Jensen93} 
Jensen I, 
Critical behavior of the pair contact process,
1993 Phys. Rev. Lett. {\bf 70}, 1465.

\bibitem{JensenDickman93} 
Jensen I and Dickman R, 
\textit{Nonequilibrium phase transitions in systems with infinitely many absorbing states},
1993 Phys. Rev. E {\bf 48}, 1710.

\bibitem{MGDL96} 
Mu{\~n}oz M A, Grinstein G, Dickman R, and Livi R, 
\textit{Critical behavior of systems with many absorbing states},
1996 Phys. Rev. Lett. {\bf 76}, 451.

\bibitem{MGD98} 
Mu{\~n}oz M A, Grinstein G, and Dickman R
\textit{Phase structure of systems with infinite numbers of absorbing states},
1998 J. Stat. Phys. {\bf 91}, 541.

\bibitem{Wijland02} 
F. van Wijland, 
\textit{Universality class of nonequilibrium phase transitions with infinitely many absorbing states},
2002 Phys. Rev. Lett. {89}, 190602.

\bibitem{Lubeck02} 
L\"ubeck S and Willmann R D, 
\textit{Universal scaling behaviour of directed percolation and the pair contact process in an external field},
2002 J. Phys. A {\bf 35}, 10205.

\bibitem{LopezMunoz97} 
L{\'o}pez C and Mu{\~n}oz M A, 
\textit{Numerical analysis of a Langevin equation for systems with infinite absorbing states},
1997 Phys. Rev. E {\bf 56}, 4864.

\bibitem{JimenezHinrichsen03}  
Jim\'enez-Dalmaroni A and Hinrichsen H,
\textit{Epidemic processes with immunization},
2003 Phys. Rev. E {\bf 68}, 036103 [cond-mat/0304113].

\bibitem{GrassbergerTorre79} 
Grassberger P and de la Torre A, 
\textit{Reggeon field theory (Schl{\"o}gls first model) on a lattice -- Monte-Carlo calculations
of critical behavior},
1979 Ann. Phys. {\bf 122}, 373.

\bibitem{Janssen03}
Janssen H K, 
\textit{Survival and Percolation Probabilities in the Field Theory of Growth Models}, 
submitted [cond-mat/0304631].

\bibitem{grassberger2003} 
Grassberger P, 
\textit{Critical percolation in high dimensions},
2003 Phys. Rev. E {\bf 67}, 036101.

\bibitem{JanssenLog}
Janssen H K and Stenull O,
\textit{Logarithmic Corrections in Directed Percolation},
2004 Phys. Rev. E  {\bf 69}, 016125.

\bibitem{Lubeck04} 
L{\"u}beck S and Willmann R D, 
\textit{Universal scaling behavior of directed percolation around the upper critical dimension},
2004 J. Stat. Phys {\bf 115}, 1231 [cond-mat/0401395].

\bibitem{munoz} 
Mu\~noz M A, Grinstein G and Tu Y,
\textit{Survival probability and field theory in systems with absorbing states},
1997 Phys. Rev. E {\bf 56}, 5101.

\bibitem{RW}
Spitzer F, \textit{Principles of Random Walk}, Springer-Verlag,
New~York, second-edition, (1976).

\bibitem{FHL01} 
Fr\"ojdh P, Howard M, and Lauritsen K B,
\textit{Directed percolation and other systems with absorbing states: Impact of boundaries},
2001 Int. J. Mod. Phys. {\bf 15}, 1761, and references therein.

\bibitem{voigt_ziff} 
Voigt C A, Ziff R M, 
\textit{Epidemic analysis of the second-order transition in the Ziff-Gulari-Barshad surface-reaction model},
1997 Phys. Rev. E {\bf 56}, R6241. 

\bibitem{Jensen92} 
Jensen I, 
\textit{Critical behavior of the 3-dimensional contact process},
1992 Phys. Rev. A {\bf 45}, R563.

\bibitem{Jensen99} 
Jensen I, 
\textit{Low-density series expansions for directed percolation: 
I. A new efficient algorithm with applications to the square lattice},
1999 J. Phys. A {\bf 32}, 5233.

\bibitem{Grassberger96}  
Grassberger P and Zhang Y C, 
\textit{''Self-organized'' formulation of standard percolation phenomena},
1996 Physica A {\bf 224}, 169.

\bibitem{reweight} Dickman R, \textit{Reweighting in nonequilibrium
  simulations}, 1999 Phys. Rev. E {\bf 60}, R2441.

\bibitem{Janssen81} 
Janssen H K, 
\textit{On the non-equilibrium phase transition in reaction-diffusion systems with an absorbing stationary state},
1981 Z. Phys. B {\bf 42}, 151.

\bibitem{ziff} Lorenz C D and Ziff R M, \textit{Precise determination
  of the bond percolation thresholds and finite-size scaling
  corrections for the sc, fcc, and bcc lattices}, 1998 Phys. Rev. E
  {\bf 57}, 230.

\bibitem{Grassberger92} Grassberger P, \textit{Numerical studies of
  critical percolation in three dimensions}, 1992 J. Phys. A {\bf 25}, 5867.
  
\bibitem{munoz2}  
Mu\~noz M A, Dickman R, Vespignani A and Zapperi S, 
\textit{Avalanche and spreading exponents in systems with absorbing states},
1999 Phys. Rev. E {\bf 59}, 6175.

\bibitem{bunde} Bunde A and Havlin S, in \textit{Fractals and
  Disordered Systems}, ed.~by Bunde A and Havlin S, Springer-Verlag,
  Heidelberg, (1991).

\end{thebibliography}
\end{document}